\documentclass[aps,prb,reprint,groupedaddress,showpacs,amsmath,amssymb,twocolumn,floatfix,preview]{revtex4-1}
\usepackage{graphicx}
\usepackage{bm}
\usepackage{color}

\newcommand{\Rd}{$R_{drag}$}

\begin{document}

\title{Negative Coulomb Drag in Double Bilayer Graphene}

\author{J.I.A. Li$^{1}$}
\author{T. Taniguchi$^{2}$}
\author{K. Watanabe$^{2}$}
\author{J. Hone$^{3}$}
\author{A. Levchenko$^{4}$}
\author{C.R. Dean$^{1}$}

\affiliation{$^{1}$Department of Physics, Columbia University, New York, NY, USA}
\affiliation{$^{2}$National Institute for Materials Science, 1-1 Namiki, Tsukuba, Japan}
\affiliation{$^{3}$Department of Mechanical Engineering, Columbia University, New York, NY, USA}
\affiliation{$^{4}$Department of Physics, University of Wisconsin-Madison, Madison, Wisconsin 53706, USA}

\date{\today}


\maketitle

\textbf{Coulomb drag \cite{Lev.15} between parallel quantum wells provides a uniquely sensitive measurement of electron correlations since the drag response depends on interactions only. \cite{Sol.89,Gra.91,Sivan.92,Eis.14} Recently it has been demonstrated that a new regime of strong interactions can be accessed for devices consisting of two monlolayer graphene (MLG) crystals, separated by few layer hexagonal boron-nitride.\cite{Dean.10,Gor.13,Tutuc11,Tutuc12,Tit.13}  Here we report measurement of Coulomb drag in a double bilayer graphene (BLG) stucture,  where the interaction potential is anticipated to be yet further enhanced compared to MLG. \cite{Sarma.11} At low temperatures and intermediate densities a new drag response with inverse sign is observed, distinct from the momentum and energy drag mechanisms previously reported in double MLG.  We demonstrate that by varying the device aspect ratio the negative drag component can be suppressed and a response showing excellent agreement with the density and temperature dependance predicted for momentum drag in double BLG is found. \cite{Hwa.11,Lux.12} Our results pave the way for pursuit of emergent phases in strongly interacting bilayers, such as the exciton condensate.}

Measurements \cite{Gor.13,Tutuc11,Tutuc12,Tit.13} of Coulomb drag in double well structures consisting of two monolayer graphene (MLG) flakes, separated by few-layer hexagonal boron-nitride (hBN), have recently received significant interest. \cite{Hwa.11,Lux.12,Kat.11,Per.11,Nar.12,Sch.12,Am.12,Princ.12,Son.12,Schutt.13,Son.13} In addition to the unique dispersion of the graphene bandstructure, advancements in the mechanical assembly of 2D materials make it possible to reduce the interlayer well distance to only a few atomic lengths, while preserving high mobility.  Moreover, the ambipolar nature of graphene allows independent control over the carrier type and density in each layer with simple electrostatic gating.  This provides experimental access to the regime of strong interactions where new phases of matter, such as the superfluid exciton condensate, are expected to emerge. \cite{Kha.08,Min.08,Loz.12,Per.13,Zar.14}

\begin{figure}
\includegraphics[width=.85\linewidth]{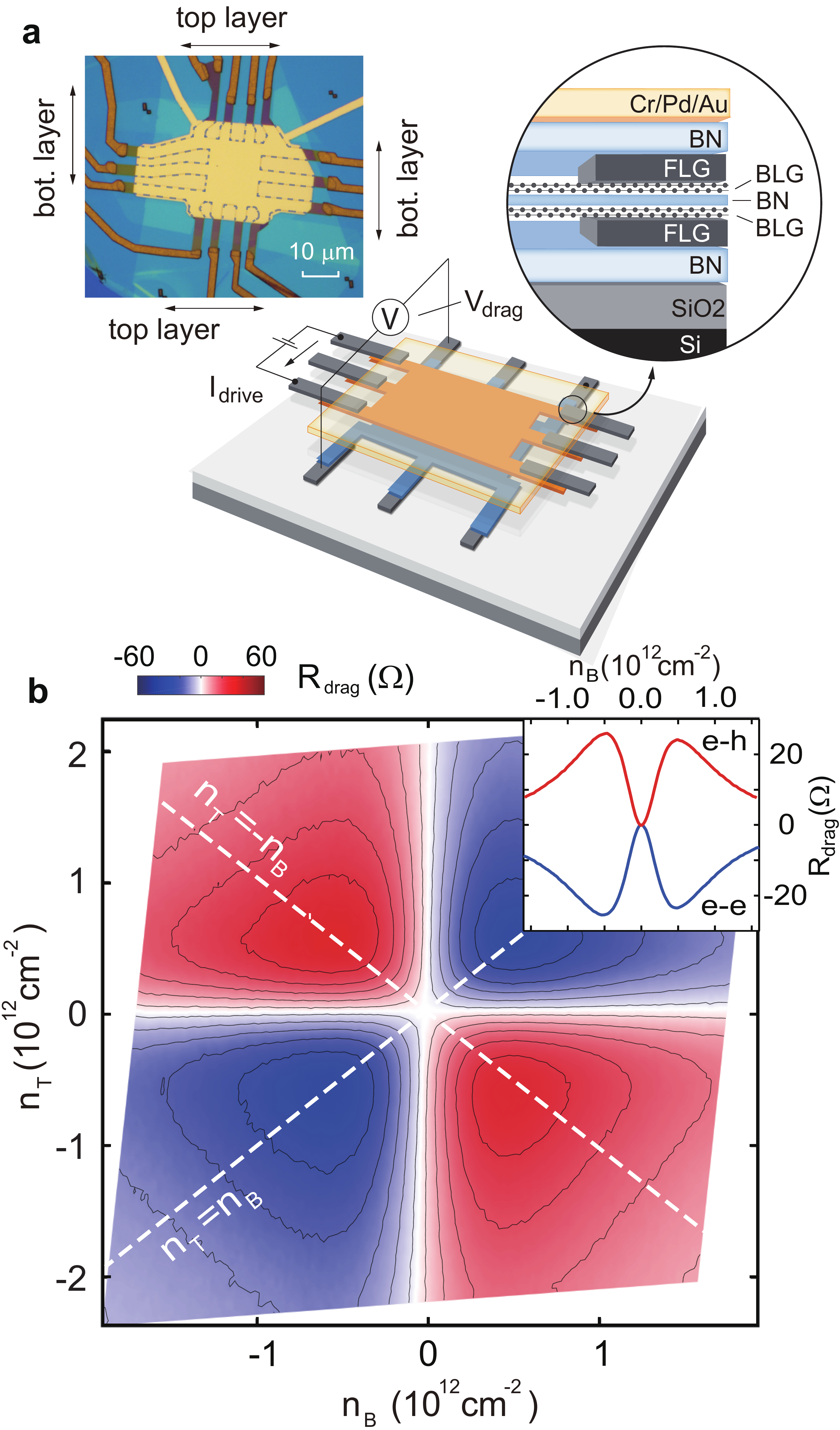}
\caption{\label{fig1}{\bf{Coulomb drag.}}  (a) Schematic of a double-bilayer graphene device and local Coulomb drag measurement. Left inset, optical image of a double-bilayer graphene device. Right inset, cross section of the bilayer graphene-hBN heterostructure. (b) $R_{drag}$ as a function of $n_T$ and $n_B$ at $300$K from the local drag measurement. The solid curves are isolevels. Inset, The behavior of $R_{drag}$ at $300$K along match density lines, $n_T = n_B$ (e-e) and $n_T = -n_B$ (e-h).}
\end{figure} 

Drag experiments in double MLG \cite{Gor.13,Tutuc11,Tutuc12,Tit.13} have indeed revealed a rich complexity of new behaviors, including a low density response at both zero and finite field driven by energy coupling mechanisms, \cite{Son.12,Schutt.13,Son.13} and a high density scaling not captured by existing theories. The precise relationship of these observations to the MLG bandstructure is the subject of ongoing  studies. In a parallel vein, owing to the different single particle energy spectrum and density of states in bilayer graphene (BLG), significant variation in the drag coefficient is expected for double quantum wells consisting of two BLG layers. \cite{Hwa.11,Lux.12}  Moreover, further enhancement of the interaction strength compared with MLG is anticipated, which could for example stabilize the condensate phase at higher temperatures.  

\begin{figure}
\includegraphics[width=1\linewidth]{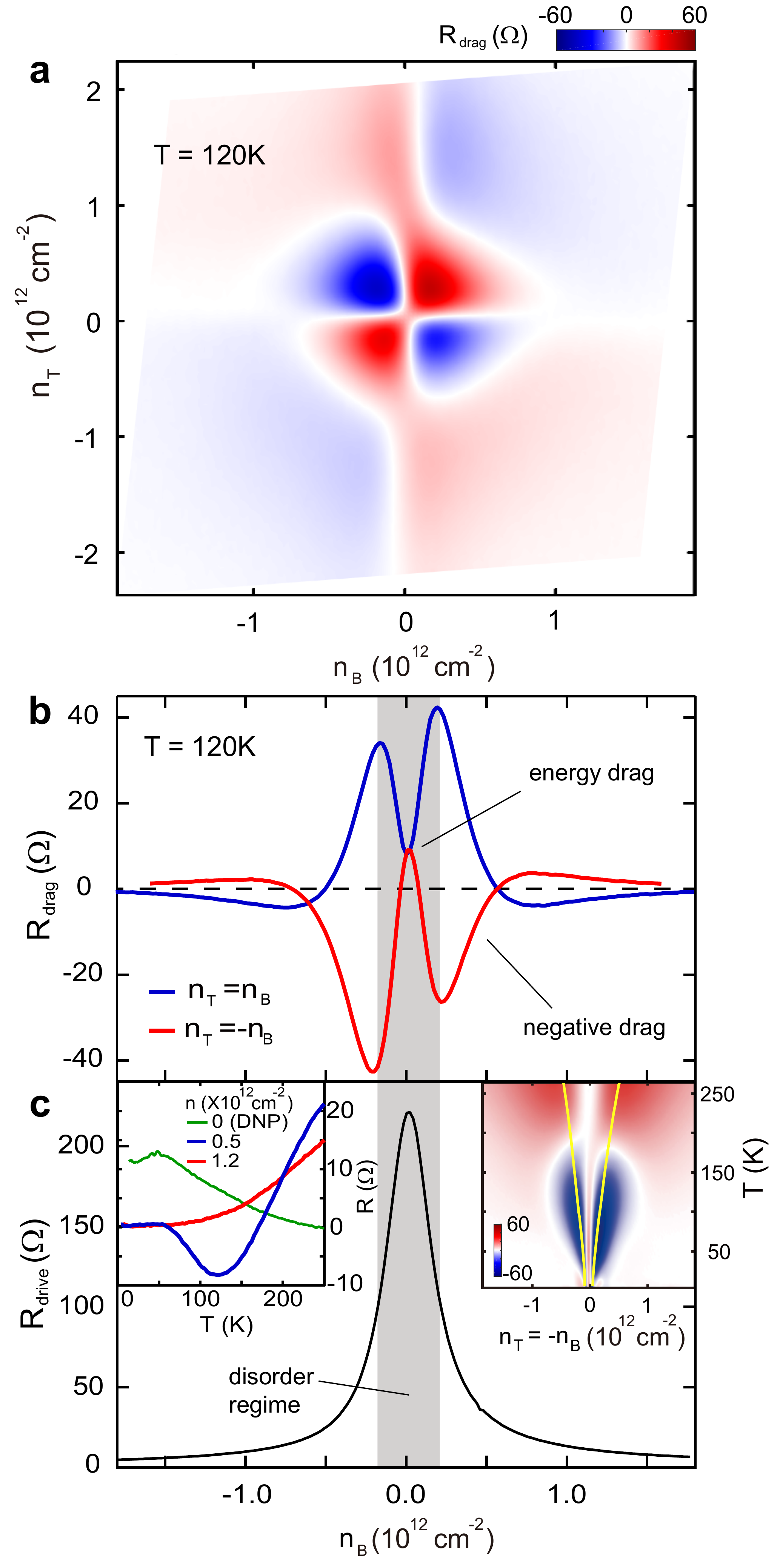}
\caption{\label{fig2} {\bf{Negative drag.}} (a) \Rd\ as a function of top and bottom layer density, $n_T$ and $n_B$, respectively, at $120$K.  (b)  \Rd\ along the matched density line, $n_T=\pm n_B$. (c) Drive layer resistivity along the same density line. The approximate charge puddle regime, defined by the full width half-maximum (FWHM) of the resistivity peak, is shown as the grey shaded area. Left inset shows the temperature dependence of \Rd\ at select densities along the $n_T=-n_B$. Right inset shows  \Rd\ under the equal density condition, $n_T = -n_B$, with varying temperature. The yellow solid line marks the FWHM of the drive layer resistivity peak with varying temperature.}
\end{figure}

\begin{figure*}
\includegraphics[width=1\linewidth]{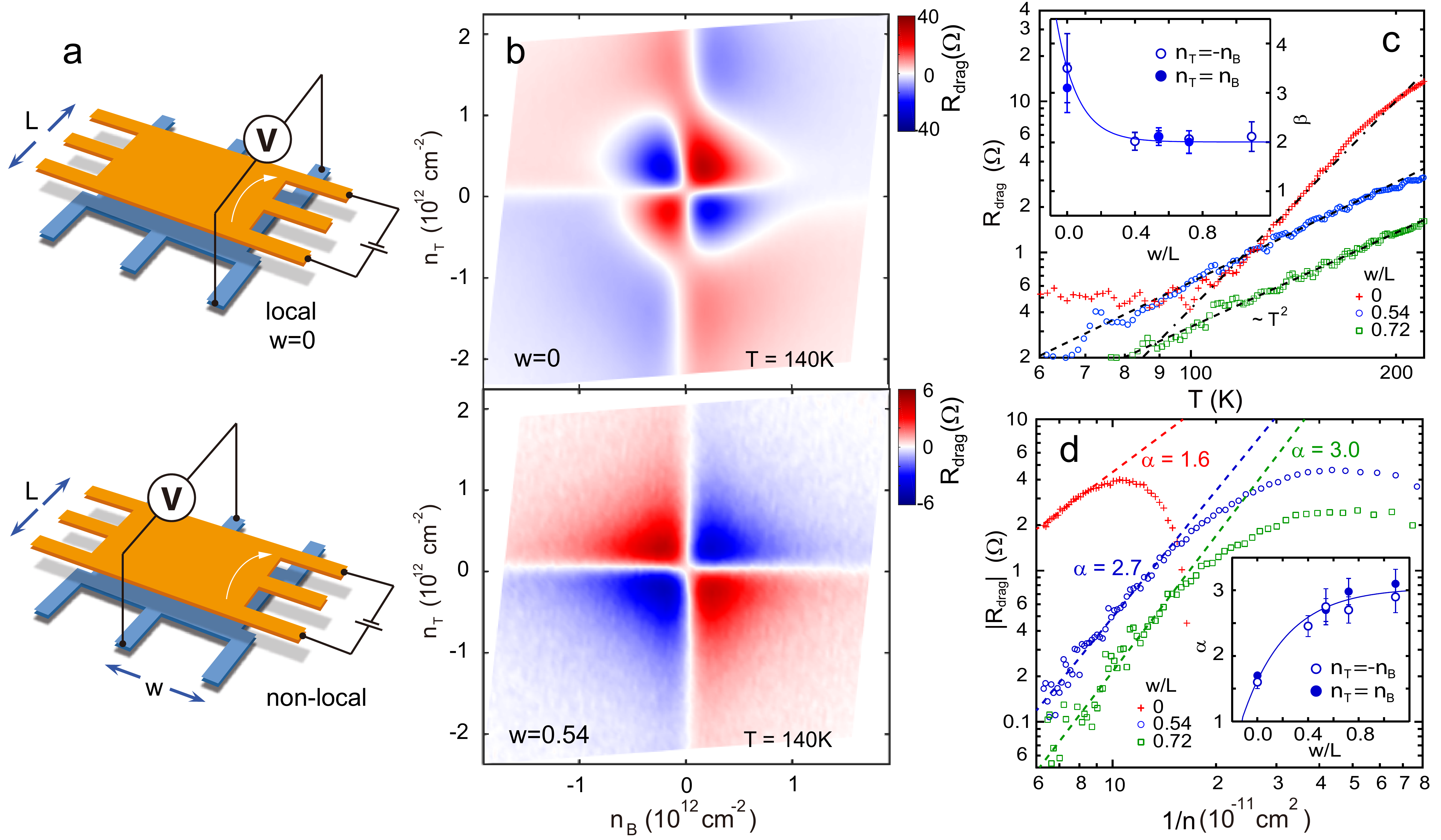}
\caption{\label{fig3} {\bf{Local and non-local drag.}} (a) shows schematic of the local and nonlocal drag measurement. (b) Upper panel, density dependence of \Rd\ from a local geometry measured at 140K.  Lower panel, density dependence of \Rd\ from a nonlocal geometry measured at 140K.  (c)  Temperature dependence of \Rd\ under equal density condition $n_T = -n_B = 7 \times 10^{11}$ cm$^{-2}$, from local and nonlocal measurements. The black dashed line corresponds to the expected $T^2$ dependence. The inset plots the value of $\beta$ versus geometric factor $w/L$, where $\beta$ is obtained by fitting \Rd\ with a power law temperature dependence. (d) \Rd\ versus inverse density in the matched density regime, $n_T = n_B$, from the local and nonlocal measurement at $T = 150$ K. The dash lines are fits to \Rd\ with a power law density dependence. The fit coefficient $\alpha$ is plotted in the inset against geometric factor $w/L$.}
\end{figure*}

Here we report Coulomb drag measurement in double BLG systems,  with interlayer hBN spacers varying from approximately 5~nm to 12~nm (see methods). In a typical drag measurement (Fig.~1a) current, $I_{drive}$, is applied through two corner leads of the drive BLG layer, and the resulting voltage, $V_{drag}$, is measured from corner leads of the drag BLG layer. Fig. 1b shows an example of the drag resistance,  defined by the relation $R_{drag} = V_{drag}/I_{drive}$, plotted as a function of the top and bottom layer densities, $n_T$ and, $n_B$, respectively, acquired at $T=300$~K. The carrier density of each BLG layer is related to the applied gate voltages by independent measurement of the layer Hall resistivities under applied magnetic field (see SI). The density dependence exhibits a $4$ quadrant symmetry, with \Rd\ being negative (positive) when the carriers in the two BLG layers have the same (opposite) sign.  This is the expected sign relation in a momentum coupling drag picture, \cite{Son.12,Schutt.13} and we adopt the convention of referring to this as positive drag in all four quadrants.  We note that all drag responses reported are similar under switching the drive and drag layers, satisfying the expected Onsager relation.

At $T=300$~K, the isolevels of \Rd\ suggests a functional dependence of \Rd\ $= f(n_T + n_B)$, as opposed to the expected form of $f(n_T \cdot\ n_B)$. \cite{Hwa.11,Lux.12} This is consistent with the drag response reported for double MLG ~\cite{Gor.13} suggesting a similar origin for the unconventional, but so far unknown, density dependence. Inset of Fig.~1b shows the drag response along the matched density condition, $n_T \pm n_B$. \Rd\ exhibits a double-peak feature, initially diverging with decreasing density, but then diminishing in the vicinity of the charge puddle regime near the charge neutrality point (CNP).  When the Fermi energy in both BLG layers is tuned to their respective CNP [referred to as the double neutrality point (DNP)] the drag response drops to zero within our measurement resolution. 

Fig.~2a shows a plot of the the drag resistance for the same measurement configuration,  but acquired at  $T=120$ K.  At this temperature the drag unexpectedly inverts sign in all four quadrants. The inversion regime remains symmetric with \Rd\ positive (negative) when both layers contain carriers with the same (opposite) sign. Examining the response along the matched density lines at low temperature (Fig.~2b and 2c) reveals three distinct drag regimes. Along $n_T=-n_B$ the sign of the drag is expected to be positive at all densities.  Instead, the drag begins positive at high density, crosses over to negative at intermediate density, and then becomes positive again at near zero density (similar behavior is apparent along $n_B=+n_T$).  A full map of the temperature and density dependance is shown inset in Fig. 2c. The high density crossover presumably results from an interplay between the conventional momentum drag, and the new negative drag mechanisms, suggesting that these two competing contributions have different density dependences. The peak negative drag response coincides approximately with the width of the transport resistivity peak near the CNP as shown in Fig. 2c., consistent with a transition to the disorder dominated puddle regime at low density. Importantly, the crossover from positive to negative drag evolves in a nontrivial way and does not track the temperature dependence of the full width at half-maximum of the resistive peak near DNP (see solid yellow line in the temperature-density plot inset in Fig. 2c), suggesting the negative drag does not have a simple correlation with sample disorder. Finally, the finite drag response at the DNP also shows nonmonotonic behavior, increasing with decreasing temperature to a maximum value at about 50~K and then decreasing again (green line inset in Fig. 2c).  The similar magnitude and temperature dependence to that observed for the DNP response in double MLG ~\cite{Gor.13}, suggests the same energy drag mechanism as the origin of this ``zero density'' feature.  

\begin{figure}
\includegraphics[width=1\linewidth]{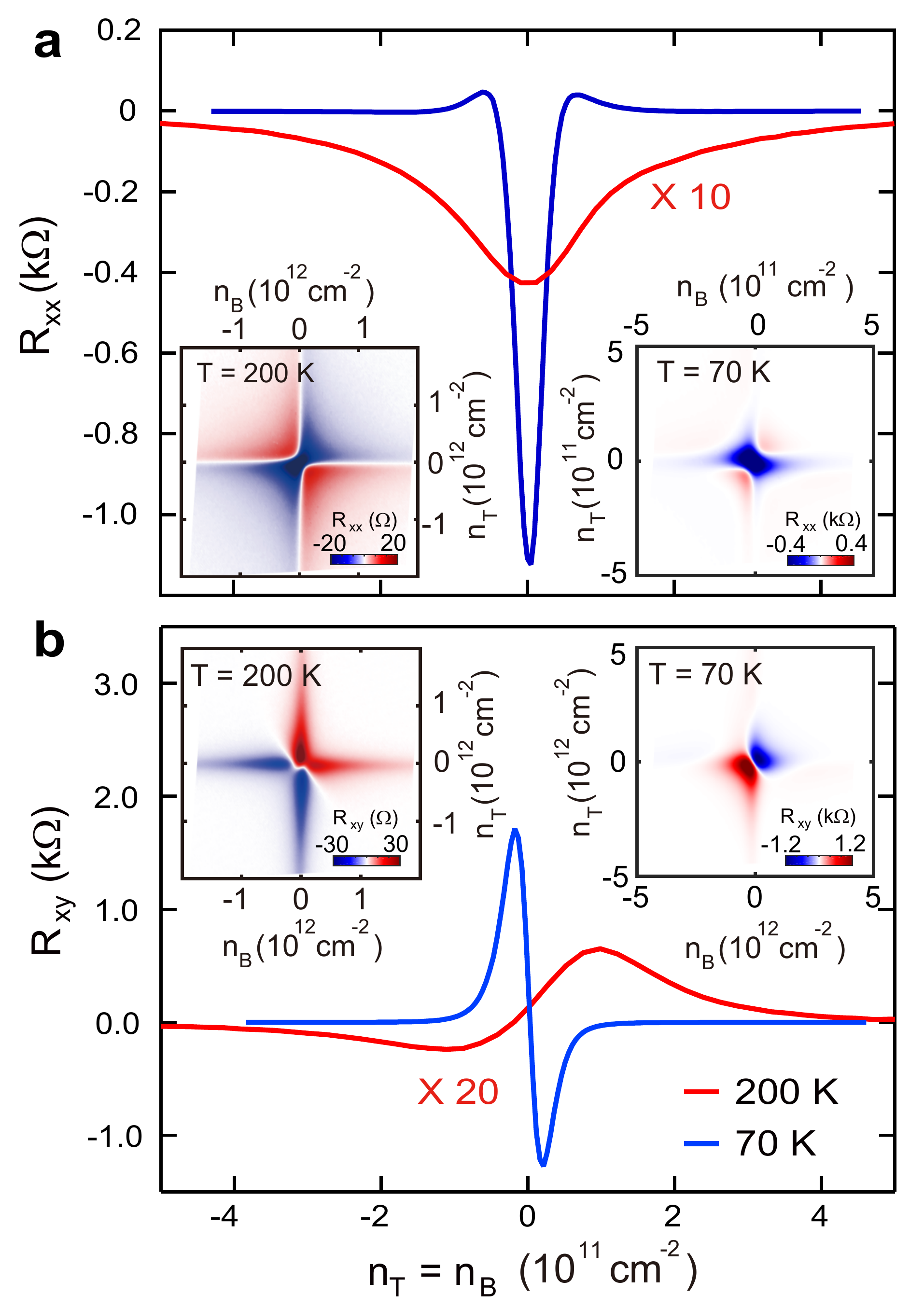}
\caption{\label{fig4} {\bf{Magnetodrag.}} a) \Rd\ grows negative with increasing magnetic field at the DNP, for measurements performed at 200K and 70K. The insets show the density dependence of \Rd\ at 200K (lower left) and 70K (lower right). b) Hall drag $R_{xy}$ measured at $T = 200$ K and $70$ K, $B = 1T$ as a function of match density, $n_T = n_B$. Inset shows the density dependence of $R_{xy}$ measured at $T = 200$ K (lower left) and $70$ K (lower right). }
\end{figure}

Figs.~3b shows the result of varying the measurement configuration. We characterize the geometry by the ratio $w/L$ where  $w$ is the lateral distance separating the current and voltage leads, and $L$ is the distance between the source and drain.   A schematic cartoon of a ``local drag'' (defined by $w= 0$)  and a ``nonlocal drag'' ($w\neq0$) measurement geometry are shown in Fig.~3a (we note that in all measurements the voltage leads remain parallel to the current leads). In the nonlocal geometry the negative drag component is suppressed (Fig. 3b), and a picture qualitatively similar to the high temperature response is fully recovered. Since interaction is mediated through long-range Coulomb scattering in the momentum transfer picture, we argue that the negative drag originates from a more local interaction between charge carriers.

In the Fermi liquid regime with drag mediated by a momentum-relaxation mechanism, the drag coefficient for double BLG, in the matched density configuration, is expected theoretically to vary with temperature, $T$, and  density, $n=\|n_{T,B}\|$, according to the scaling formula:\cite{Hwa.11,Lux.12} 
\begin{equation}
R_{drag} \propto \frac{T^{\beta}}{n^\alpha}\label{1}\\ 
\end{equation} 
with temperature and density power exponents dependent on a particular transport regime defined by 
the Fermi energy $E_F$, Fermi momentum $k_F$, interlayer separation $d$, and inverse Thomas-Fermi screening radius $k_{TF}$, respectively. For low temperatures when $T\ll E_F/(k_Fd)$ the temperature exponent is $\beta=2$ for both high ($k_Fd\gg1$) and low ($k_Fd\ll1$) density limits irrespective of interlayer coupling strength. The density exponent is $\alpha=3$ except for the low density $k_Fd\ll1$ and weak coupling $k_{TF}d\gg1$ limit when $\alpha=2$. The high temperature regime, $T\gg E_F/(k_Fd)$, is only meaningful for the high density limit $k_Fd\gg1$. In this case $\beta=1$  and $\alpha=5/2$ for any value of $k_{TF}d$. Since screening radius in a BLG is independent of density, we estimate our samples to be in a strong coupling regime with $k_{TF}d\sim0.6$ and always at temperatures satisfying $T\ll E_F/(k_Fd)$. We then expect for high densities $k_Fd\gg1$ 
\begin{equation}
R_{drag}\simeq\frac{h}{e^2}\left(\frac{T}{E_F}\right)^2\left(\frac{k_{TF}}{k_F}\right)^2\ln\left(\frac{1}{k_{TF}d}\right),
\end{equation}
whereas at low densities $k_Fd\ll1$ the logarithm in the above formula should be replaced by $[\ln(1+k_F/4k_{TF})-k_F/(4k_{TF}+k_F)]$. Since in a BLG $E_{F}\propto n$ and $k_F\propto\sqrt{n}$ then at a strong coupling we generically expect $R_{drag}\propto T^2/n^3$ for momentum-relaxation mechanism of drag. 

In Fig.~3c we compare the temperature dependence of \Rd\ in the equal density regime $n_T = -n_B$, from the local and nonlocal geometry. In the nonlocal geometry, the response appears to well fit a power law over large temperature range whereas the local drag response displays significant deviation. We interpret this to be a consequence of the competing mechanisms of the positive and negative drag components, with the relative contributions apparently varying with temperature.   In the inset, the power law coefficient $\beta$ is plotted against the geometric factor $w/L$.  The contribution from the negative component is increasingly suppressed as the measurement geometry is made more nonlocal, and the power converges to the expected value of $\beta = 2$ within the measurement uncertainty [the same result is observed for $n_T = n_B$ (see SI)].  Fig.~3d shows the density dependence of \Rd\ in the equal density regime at $T=150$~K, for different measurement geometries.  With increasing nonlocal geometry, the density dependence of \Rd\ converges to the expected $1/n^{\alpha}$ dependance, with $\alpha = 3$ (inset in Fig.~3d).  Both the temperature and density response suggests that by measuring in the nonlocal geometry we are able to isolate the momentum coupling component of the drag response, and moreover we find good quantitative agreement  with the theoretically calculated temperature and density dependence for double BLG. \cite{Hwa.11,Lux.12}

Finally, we examine the drag response in the presence of a magnetic field. Near the DNP, \Rd\ is shown to be negative in a small magnetic field, and grows in amplitude with increasing $B$ field. This behavior is consistent with previous result from MLG, \cite{Gor.13} originating from an energy driven Nernst effect.\cite{Son.12,Tit.13}  At $B = 1$ T, the density dependence of \Rd\ measured at $ T = 200$ K and $70$ K are shown in the bottom left and upper right corner of Fig.~4a. At $T = 200$ K, \Rd\ displays the four quadrant symmetry consistent with momentum drag.  At $70$ K, \Rd\ changes sign away from the DNP. The sign inversion is particularly clear in the $n_T = n_B$ (e-e and h-h) quadrants, contrasted by the strong negative peak at the DNP. Simultaneous Hall drag is measured at $200$ K and $70$ K, Fig.~4b and c, showing that the amplitude of $R_{xy}$ increases with increasing field. Hall drag is expected to be zero in a pure momentum transfer picture and a nonzero Hall drag response can be explained by the field induced coupling between the momentum and energy transfer modes. \cite{Son.13,Tit.13} Most interestingly, in the regime of negative drag, Hall response changes sign from $200$ K to $70$ K, further indicating that the negative drag has a different origin compared to the momentum and energy transfer mechanism. 

At present the origin of the negative drag is not known. Because of its appearance in all four density quadrants we do not consider this to be related to formation of indirect excitons between the layers. The suppression in non-local geometry suggests the negative drag results from a shorter relaxation mechanism than can be attributed to a momentum coupling picture. There is also no obvious reason to believe there is a relation to the mechanism of negative drag reported for 1D-1D systems. \cite{Yamamoto,Gervais} Finally, we note that negative drag was observed in all BLG devices studied (with interlayer distances spanning 5~nm-12~nm), but that a similar feature has not been reported for MLG, suggesting a possible relation to the dispersion relation which is quadratic in BLG compared to linear for MLG.  Further theoretical and experimental work will be needed to resolve this.

In summary, Coulomb drag measurement is reported for the first time in a double well consisting of two graphene bilayers.  At low temperature and intermediate density, a negative drag is observed with sign opposite to that expected in a simple momentum coupling regime.  We find that the negative drag response can be suppressed using a nonlocal measurement geometry, and that the temperature and density dependence of \Rd\ from nonlocal measurement matches well with theory for the momentum transfer drag.\cite{Hwa.11,Lux.12} In a low magnetic field, Hall drag and magnetodrag observed at high temperature are consistent with the energy driven mechanism observed in double MLG,\cite{Tit.13} whereas in the negative drag regime, Hall drag changes sign. Finally, we note that the capability to achieve good electrical contact to a double BLG structure, and to isolate the momentum driven drag component in a nonlocal geometry, over wide density range, makes it feasible to look for the excitonic condensate phase, possibly with smaller interlayer separation, and at lower temperature.

\section*{Methods}
The double BLG quantum wells are fabricated from exfoliated crystals, using the van der Waals assembly technique described previously \cite{Lei.13} . In our devices, each BLG is contacted with two pieces of few layer graphite (typical thickness is  $5-10$~nm) serving as electrical leads.  The entire heterostructure, consisting of 9 layers of exfoliated 2D materials, is assembled on an oxidized, doped Si substrate, and then etched into a crossed hall bar geometry (Fig. 1). Inset in Fig. 1a, shows a schematic cross section of the full layer structure in the region where the top and bottom layer graphite leads overlap.  The carrier density can be tuned independently in the top and bottom BLG layers by biasing the top evaporated metal, and bottom, doped Si, gate electrodes, respectively. The graphite leads allow us to tune the BLG layers to opposite carrier type, while maintaining good electrical contact to each layer (in double BLG, leads defined by etching, such as in previous studies of double MLG structures ~\cite{Tutuc11,Tutuc12,Gor.13,Tit.13} develop a band gap under transverse magnetic field and become highly resistive). Further details of the device fabrication including the effect of introducing graphite leads can be found in the supplementary information (SI). Similar negative drag behavior at zero field, magnetodrag and Hall drag responses are observed in all BLG devices studied (with interlayer distances spanning 5~nm-12~nm). Data shown in Fig.~1 to Fig.~3 are from a double BLG device with interlayer separation of $5.2$ nm and Fig.~4 is from a device with interlayer separation of $10.2$ nm.

\begin{acknowledgments}
This work was supported by the National Science Foundation (DMR-1507788) and by the David and Lucille Packard Foundation. A.L. acknowledges financial support by NSF grant no. DMR-1606517 and ECCS-1560732, and  by the Office of the Vice Chancellor for Research and Graduate Education with funding from the Wisconsin Alumni Research Foundation.
\end{acknowledgments}



\section*{Competing financial interests}
The authors declare no competing financial interests.


\makeatletter
\makeatletter \renewcommand{\fnum@figure}
{\figurename~SI\thefigure}
\makeatother
\subsection{Supplementary Information:\\
``Negative Coulomb Drag in Double Bilayer Graphene''}

{\bf SI.1\,\,\, Device Assembling Sequence}\\

In a typical double layer graphene Coulomb drag device, the doped Si substrate is used as the bottom gate electrode, and a top metal electrode is evaporated on the top of the device. After shaping the device, metal side contact is made to the 2D edge of the graphene leads ~\cite{Lei.13}.  To avoid creating a short between two active layers when making side contact, leads of the top BLG need to be directly exposed to the Si substrate without the screening of the bottom graphene layer. As a result the graphene leads are biased by the bottom gate electrode, forming $p-n$ junction when the BLG layers are tuned to opposite carrier type.

\begin{figure*}
\centerline{\includegraphics[width=0.85\linewidth]{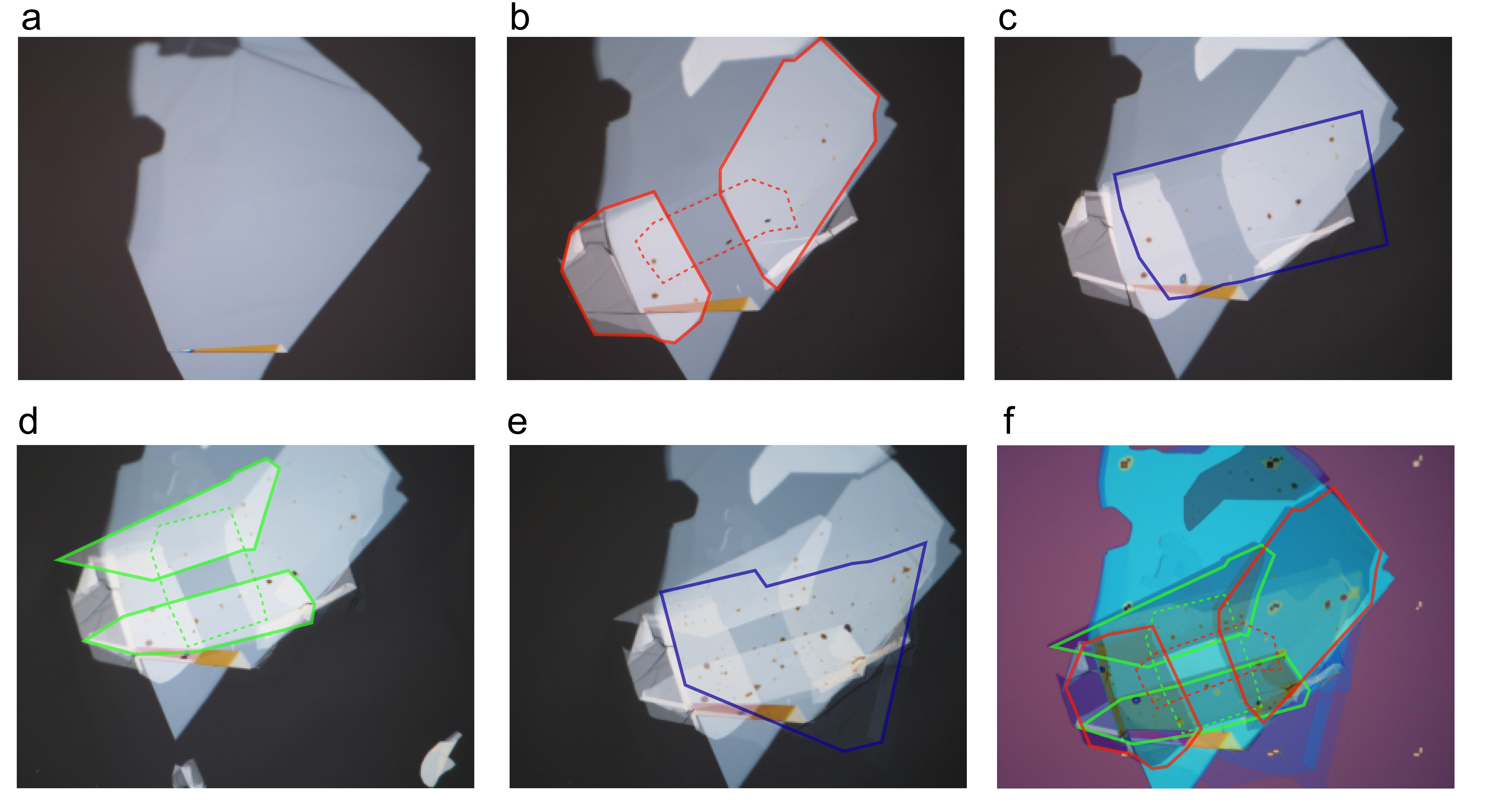}}
\caption{\label{figSI2} (Color online). (a) The top surface of the first hBN crystal is in direct contact of the polymer. Subsequent layers are picked up by this layer of hBN, therefore remaining polymer free. (b) Top active layer consists of a bilayer graphene in contact with two graphite flakes. The contour of graphene and graphite flakes are highlighted with red dashed and solid line respectively. The three flakes are picked up in three different steps. (c) A thin layer of hBN crystal, $\sim 4$ nm. (d) Bottom active layer consists of a bilayer graphene in contact with two graphite flakes. The contour of graphene and graphite flakes are highlighted with green dashed and solid line respectively. The three flakes are picked up in three different steps. (e) The bottom hBN layer that encapsulates the heterostructure. (f) The device is deposited on an oxidized Si substrate by melting the thin film of polymer. The contour of both top and bottom active layers are highlighted. The scale is set by the align mark in (f), which are separated by $50 \mu$m.   }
\end{figure*}

\begin{figure*}
\centerline{\includegraphics[width=0.8\linewidth]{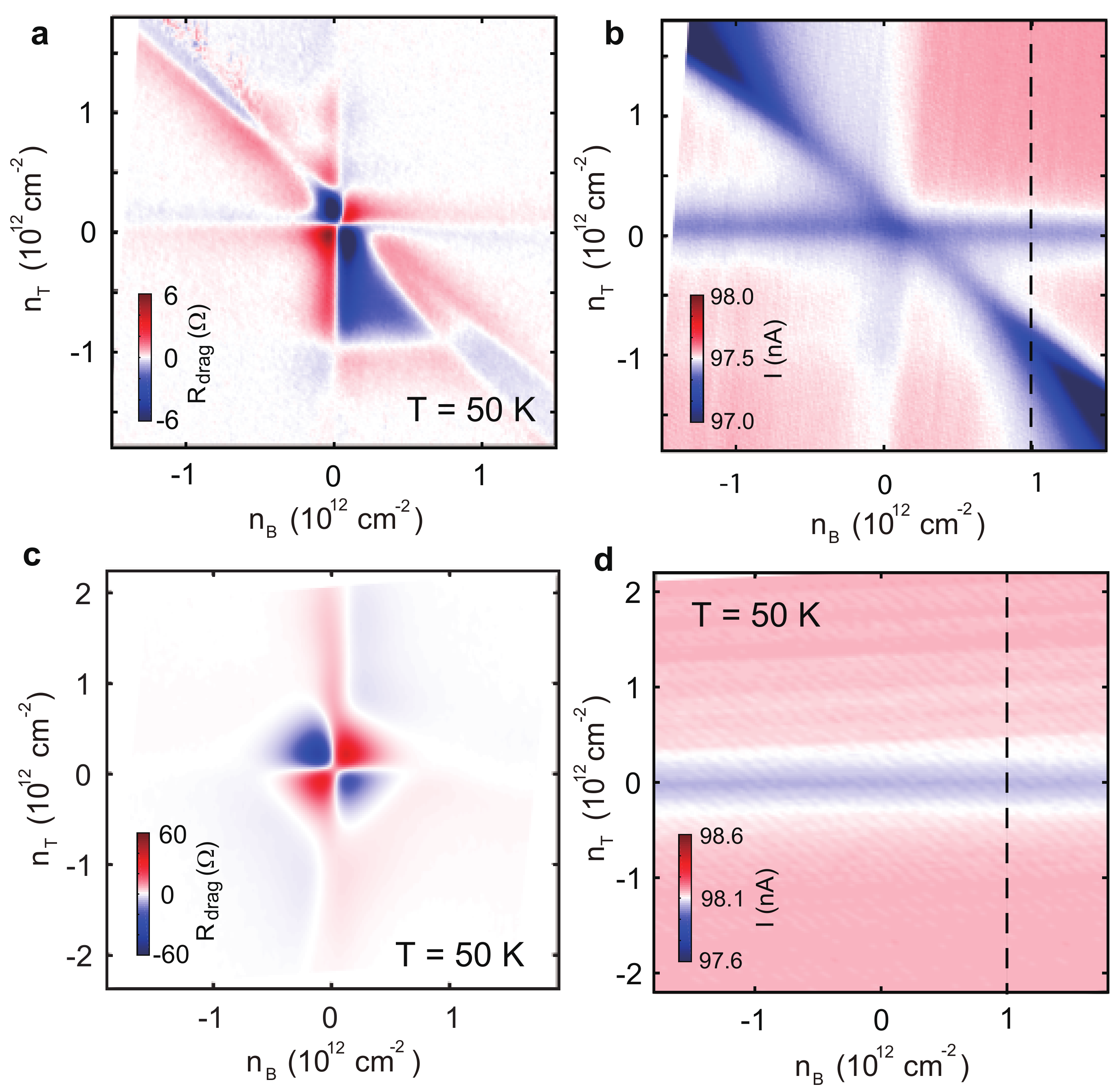}}
\caption{\label{figSI1} (Color online). (a) and (b) are drag and active layer response of a double bilayer graphene device without graphite leads at $T = 50$ K and $B = 0$ T, the interlayer separation is $d = 10$ nm. (a) Drag response as a function of top and bottom BLG density. (b) Current measured in the drive layer current in the active layer. }
\end{figure*}

\begin{figure*}
\centerline{\includegraphics[width=0.55\linewidth]{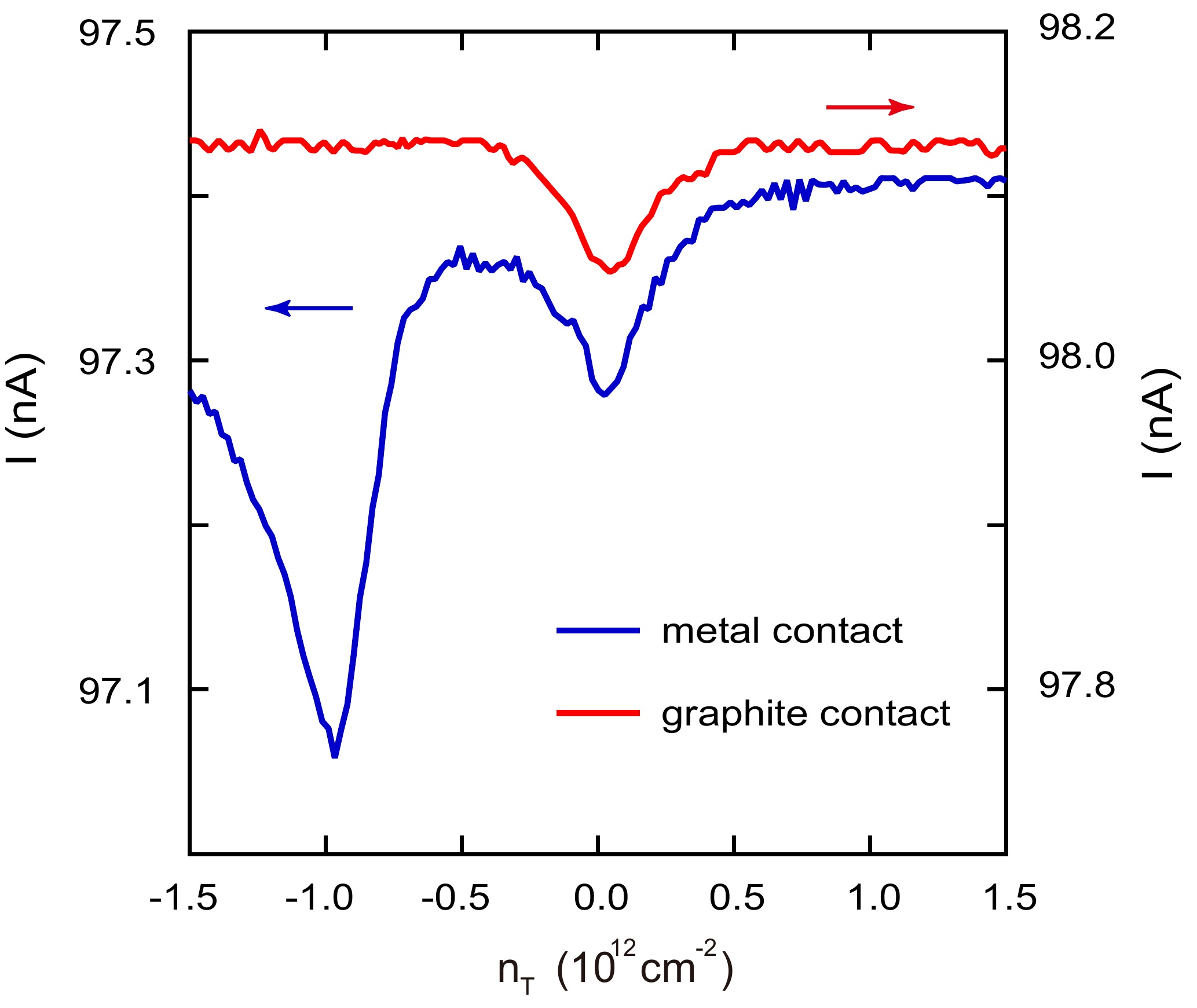}}
\caption{\label{figSI7} (Color online). Drive current measured along the dashed line shown in Fig.~SI\ref{figSI1}b and d. In devices with graphene leads, solid blue curve, an extra feature appear along the matched density line, whereas the device with graphite leads only exhibits the CNP of the graphene channel (solid red curve). }
\end{figure*}

To address these issues, we developed a new sample geometry for double layer graphene device. Each graphene layer is contacted with two pieces of few layer graphite serving as electrical leads.  The entire heterostructure, consisting of 9 layers of exfoliated 2D materials, is assembled using the van der Waals polymer free technique. All the layers, save for the top surface of the first hBN crystal, never come into contact with polymers and remains clean. Fig.~SI\ref{figSI2}a to e are false color images showing how each layer of the heterostructure is assembled affixed on the thin film of poly-propylene carbonate (PPC) and the elastomer stamp (polydimethyl siloxane, PDMS). Fig.~SI\ref{figSI2}f shows the entire heterostructure being deposited on the ozidized Si substrate, with PPC removed by vacuum annealing. The contour line indicates that the each active layer is completely shielded by the other layer from the opposite electrode.

Fig.~SI\ref{figSI1}a shows longitudinal drag response from a double BLG device with graphene leads at $T = 50$ K. Apart from the sign inversion near the double neutrality point (DNP), strong drag response are observed along the matched density regime, $n_T = -n_B$. Simultaneous current measurement reveals significant variation in drive current. The drive current measured along the dashed line in Fig.~SI\ref{figSI2}b is plotted in Fig.~SI\ref{figSI7} (solid red curve), showing a strong peak under the equal density condition where the graphene leads become resistive. Electrical contact through graphene leads also deteriorates near the charge neutrality point (CNP), $n_T = 0$ or $n_B = 0$. Strong features in the drag response and drive layer current develops in these regime,  which interferes with drag measurements. This is especially not ideal considering the possible superfluid condensation is expected to occur at low carrier density ~\cite{Per.13}. T

With the new set up, graphite leads for both active layers have constant carrier density and are immune to gate induced effect. Drag measurement at $T = 50$ K is shown in Fig.~SI\ref{figSI1}c, showing no features in the matched density regime. Simultaneous current measurement displays a weak feature corresponding to the CNP of the active layer channel (Fig.~SI\ref{figSI1}d).   The drive current along the dashed line in Fig.~SI\ref{figSI2}d is plotted in Fig.~SI\ref{figSI7} (solid red curve), indicating that good electrical contact is achieved with graphite leads.

The fabrication process includes, in sequential order, plasma shaping, metal side contact, $30$ nm of HfO$_2$ grown with ALD and metal deposition of the top gate electrode. Alternatively, the top electrode can be deposited directly on the top of the hBN crystal as the first step, then the device is shaped with plasma, and metal contact is made as the last step of the fabrication.\\

{\bf SI.2\,\,\, Drag Versus Density}\\

Fig.~SI\ref{figSI2}a shows \Rd\ as a function of the gate bias of the top and bottom electrodes. The carrier density of each BLG is weakly dependent on the gate bias of the opposite electrode, due to the partial screening of the opposite layer. Carrier density can be calculated from Hall resistance $R_{zy}$ measured in a weak magnetic field, $B = 0.2$ T,

\begin{equation}
n = \frac{B}{R_{xy}e}. \label{Eq1}\\ 
\end{equation} 
\noindent
Using this independent measurement, density of each BLG layer is calculated and \Rd\ is plotted as a function of the carrier density in Fig.~SI\ref{figSI2}.\\

\begin{figure*}
\centerline{\includegraphics[width=0.85\linewidth]{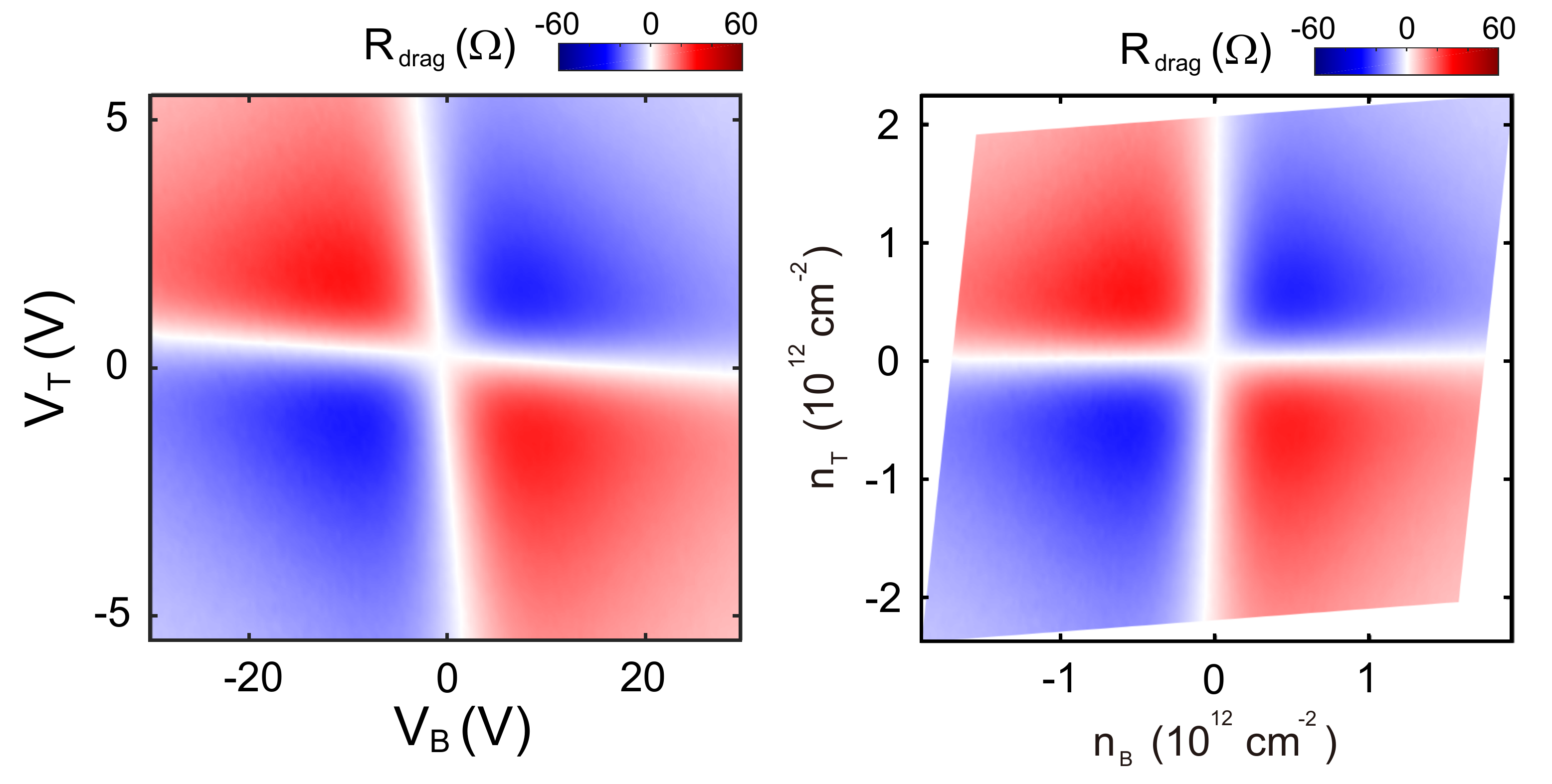}}
\caption{\label{figSI2} (Color online). (a) \Rd\ versus applied voltage on top and bottom gate electrode, $V_T$ and $V_B$. (b) \Rd\ as a function of top and bottom BLG carrier density, $n_T$ and $n_B$. }
\end{figure*}

{\bf SI.3\,\,\, Invers Density and Temperature Dependence}\\

\begin{figure*}
\centerline{\includegraphics[width=0.85\linewidth]{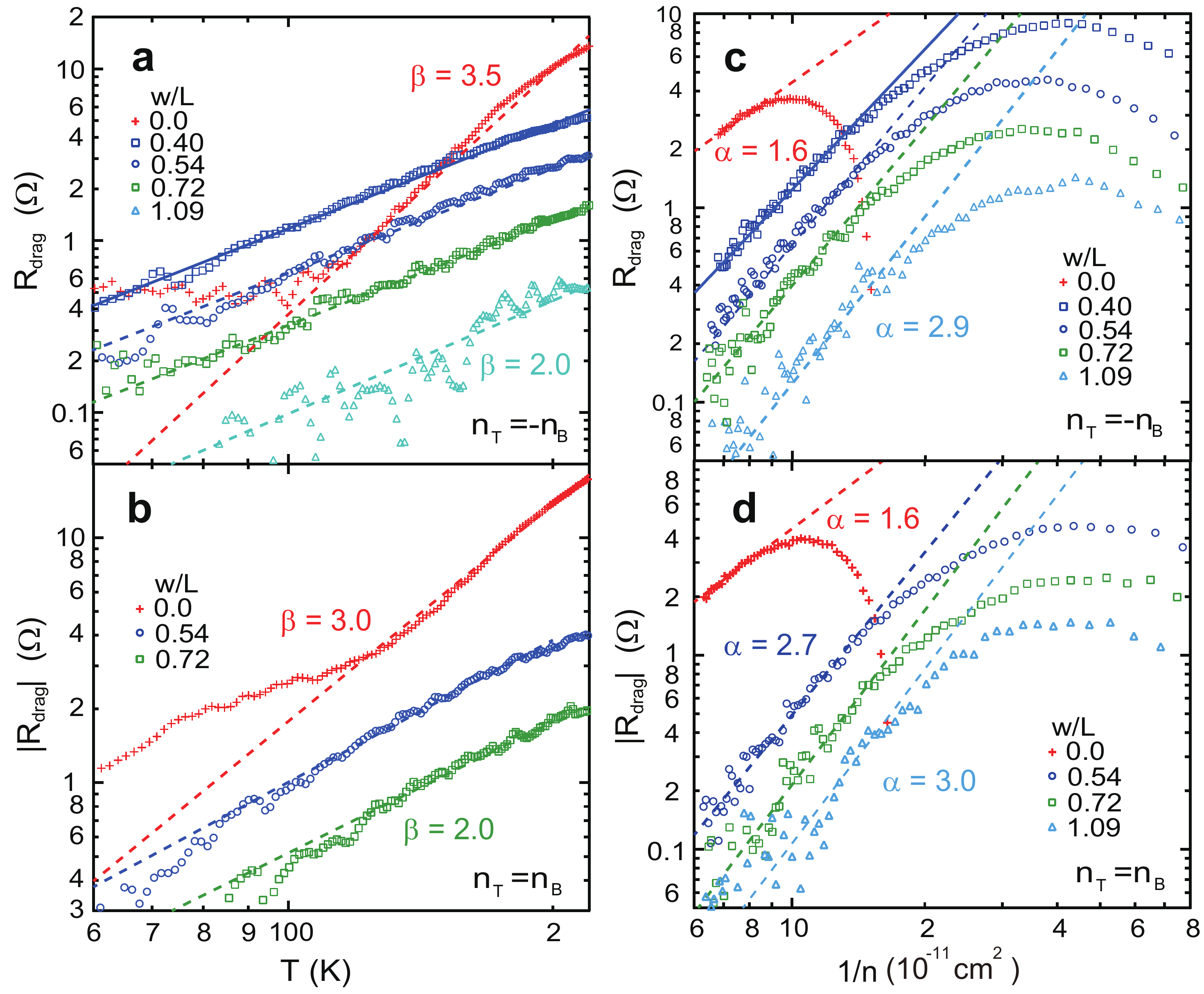}}
\caption{\label{figSI5} (Color online). (a) Temperature dependence of \Rd\ under equal density condition $n_T = -n_B = 7 \times 10^{11}$ cm$^{-2}$, from local and non-local measurements. (b) Temperature dependence of \Rd\ under equal density condition $n_T = n_B = 7 \times 10^{11}$ cm$^{-2}$, from local and non-local measurements. (c) \Rd\ versus inverse density in the matched density regime, $n_T = -n_B$, from the local and non-local measurement at $T = 150$ K. (d) \Rd\ versus inverse density in the matched density regime, $n_T = -n_B$, from the local and non-local measurement at $T = 150$ K. }
\end{figure*}

Fgi.~SI\ref{figSI5} shows the inverse density and temperature dependence of the amplitude of the drag response from both local and non-local measurement geometry. In the matched density regime, $n_T \pm\ n_B$, \Rd\ from the local measurement is a combination of the negative and positive components, whereas \Rd\ from the non-local measurement can be best described by the momentum relaxation picture ~\cite{Hwa.11}.\\

{\bf SI.4\,\,\, Geometry Dependence}\\

\begin{figure*}
\centerline{\includegraphics[width=0.85\linewidth]{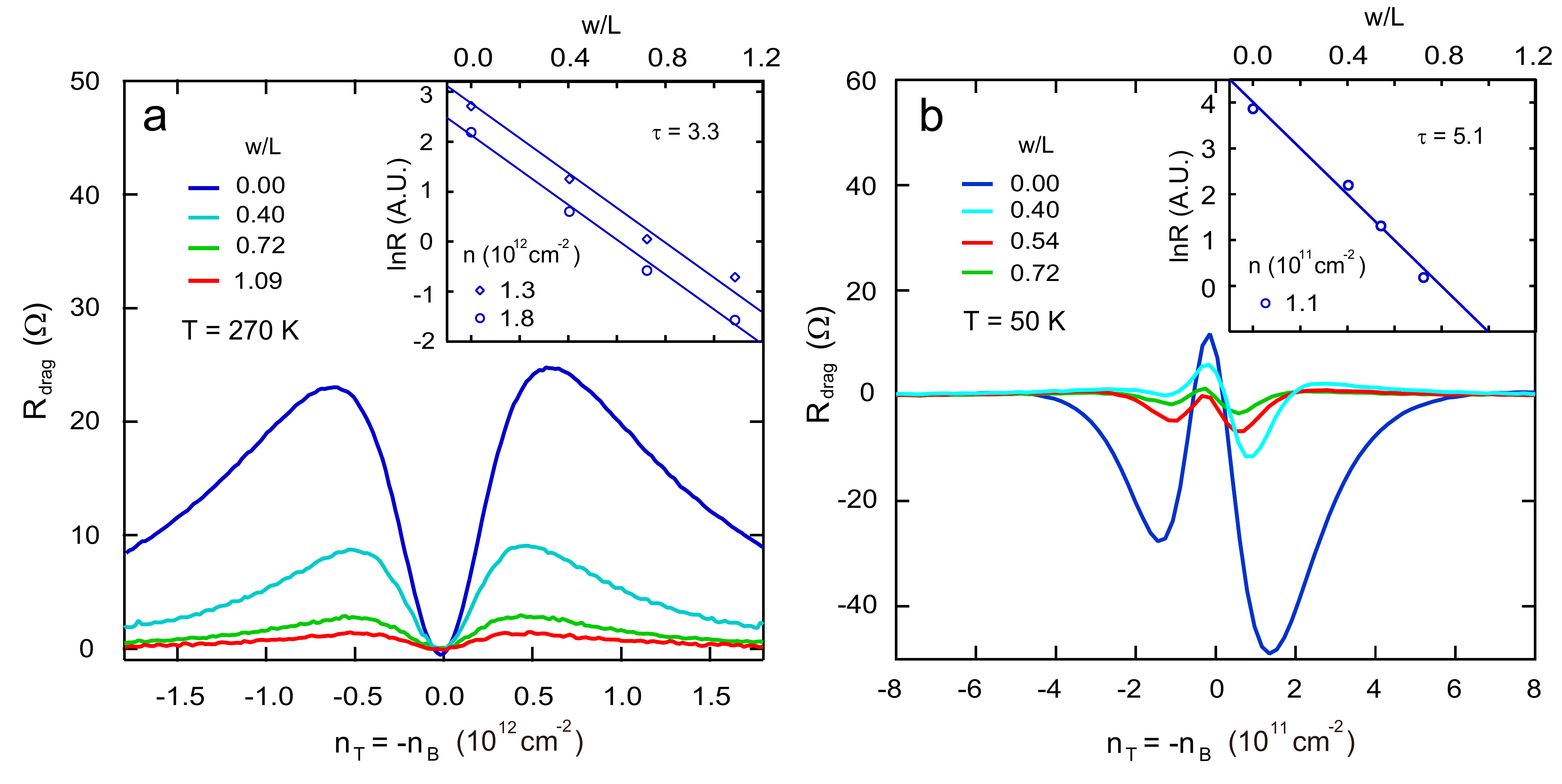}}
\caption{\label{figSI3} (Color online). (a) \Rd\ along the matched density regime for different measurement geometry at $T = 270$ K. Inset shows that for fixed density, drag response at this temperature has exponential dependence on the geometric factor $w/L$, with $\tau = 3.3$. (b) At $T = 50$ K, \Rd\ along the matched density regime for different measurement geometry show negative response at low density. Inset shows drag response in the negative drag regime as a function of the geometric facotr $w/L$.  }
\end{figure*}

Fig.~SI\ref{figSI3} shows that both the positive and negative components of the drag response exhibit exponential dependence on the geometric factor $w/L$, 

\begin{equation}
R_{drag} \propto exp(-\tau \frac{w}{L}),\label{1}\\ 
\end{equation} 
\noindent
where $w$ is the width of the device, defined as the lateral distance separting the current and voltage leads, $L$ is the length of the device, and $\tau$ is the numerical factor that characterizes the exponential decay of the drag reponse. For the negative drag, the amplitude of the signal has a steeper exponential decay compared to the positive drag and therefore larger value of $\tau$ (inset of Fig.~SI\ref{figSI3}a and b). As a result, the negative drag is suppressed in the non-local measurement geometry, and we are able to isolate the contribution from the momentum component (Fig.~3d of the main text). We note that for the non-local measurement geometry, negative drag response is only observed at low temperature, $T \leq\ 50$ K, and low density, $n \sim\ 10^{11}$ cm$^{2}$.  \\

{\bf SI.5\,\,\, The Negative and Positive Drag Components}\\

\begin{figure*}
\centerline{\includegraphics[width=0.85\linewidth]{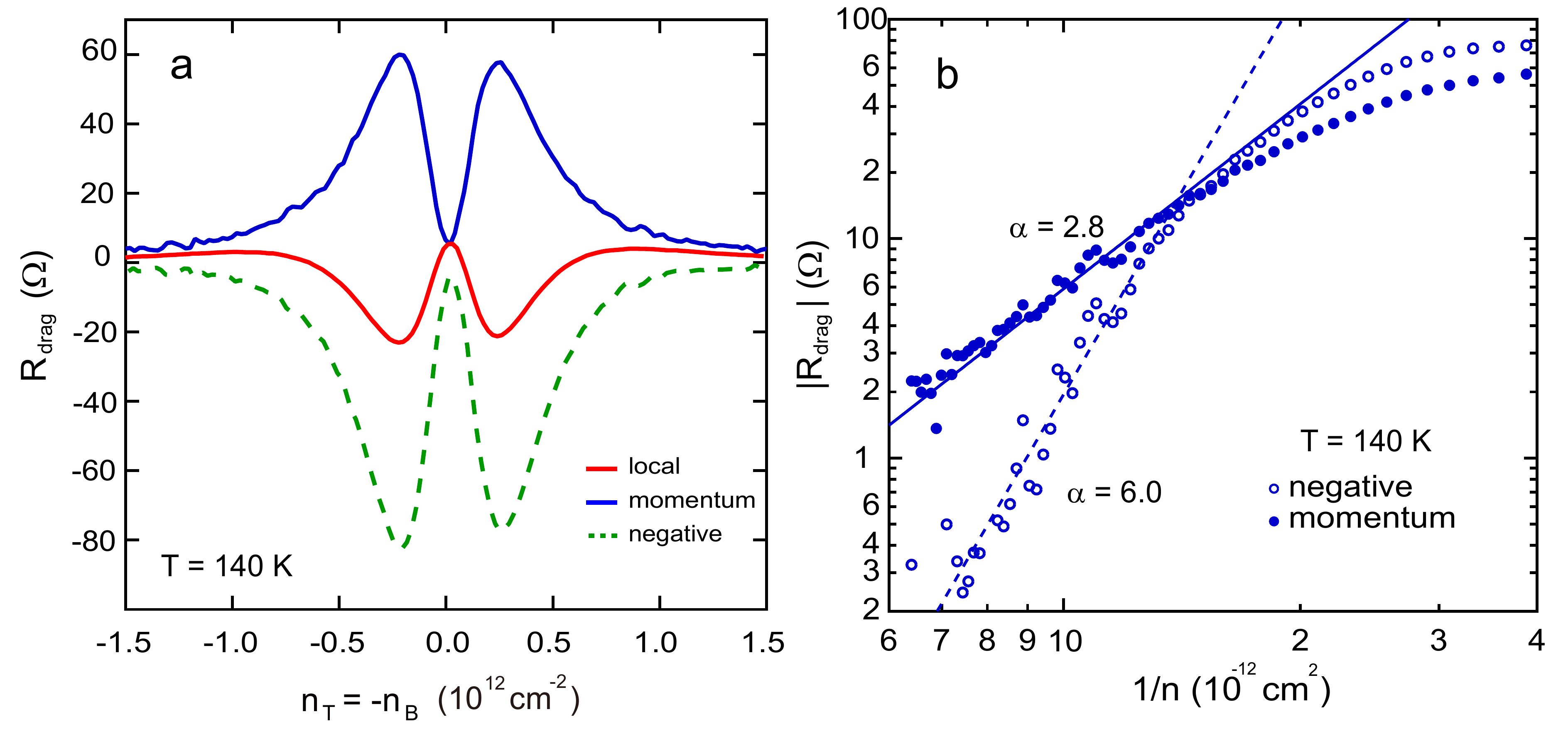}}
\caption{\label{figSI4} (Color online). (a) At $T = 150$ K, \Rd\ from the local measurement geometry in the matched density regime (red solid curve). The momentum and negative components of the drag resistance are separated and plotted as the blue solid and green dashed curves respectively.  (b) The amplitude of the positive and negative as a function of inverse density.  }
\end{figure*}

The negative drag component can be isolated from total drag response in the local measurement, allowing us to study its density dependence $R \propto 1/(n^{\alpha})$. Following the exponential behavior described in SI.3, the momentum component of the local drag resistance can be extrapolated from the non-local measurement (solid blue curve in Fig.~SI\ref{figSI4}a), and the difference compared to the total drag resistance is the negative drag response (green dashed line in Fig.~SI\ref{figSI4}a). As a function of inverse density, the negative drag component shows power law dependence with $\alpha = 6$, in contrast with $\alpha = 3$ for the momentum component. This is consistent with the observation that the negative drag dominates at intermediate density and the momentum drag is recovered in the high density regime. \\

{\bf SI.5\,\,\, Interlayer Distance}\\

\begin{figure*}
\centerline{\includegraphics[width=1\linewidth]{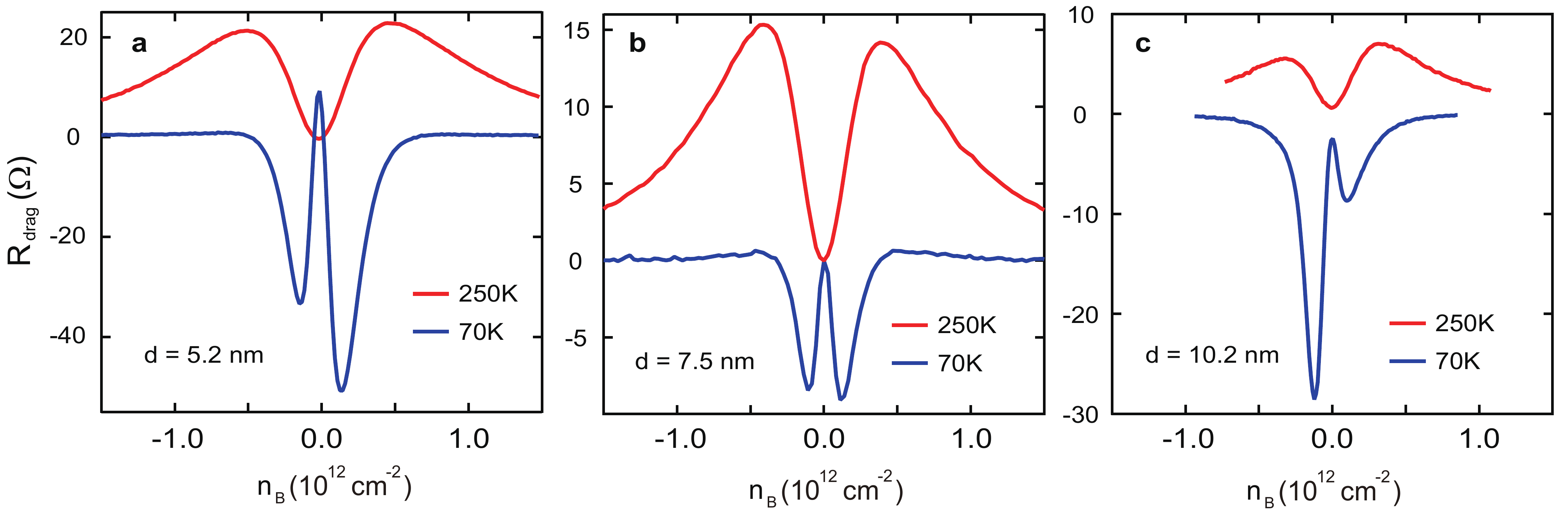}}
\caption{\label{figSI8} (Color online). Drag response measured along the matched density line, $n_T=-n_B$, for devices with different interlayer distance $d$. (a) $d = 5.2$ nm. (b) $d = 7.5$ nm. (c) $d = 10.2$ nm.   }
\end{figure*}

Negative drag behavior at zero field is observed in all BLG devices studied (with interlayer distances spanning 5~nm-12~nm). Fig.~SI\ref{figSI8} compares drag response measured at $T = 250$ and $70$ K for devices with different interlayer distance.

\end{document}